\pdfoutput=1 
\documentclass[conference]{IEEEtran}
\usepackage{cite}

\ifCLASSINFOpdf
   \usepackage[pdftex]{graphicx}
   \graphicspath{{./figures/}}
\else
\fi
\usepackage[cmex10]{amsmath}
\usepackage{amssymb}

\usepackage{multirow}	
\usepackage[caption=false,font=footnotesize]{subfig}

\begin{document}

\title{Optimization of Discrete-parameter Multiprocessor
Systems using a Novel Ergodic Interpolation
Technique}

	\author{
	\IEEEauthorblockN{Neha V. Karanjkar \  and \  Madhav P. Desai}
	\IEEEauthorblockA{Department of Electrical Engineering,\\
	Indian Institute of Technology Bombay\\
	email: \small{\texttt{\{nehak,madhav\}@ee.iitb.ac.in}}}}


\maketitle

\providecommand{\ceil}[1]{\left \lceil #1 \right \rceil }
\providecommand{\floor}[1]{\left \lfloor #1 \right \rfloor }

\begin{abstract}
Modern multi-core systems have a large number of design parameters, most of
which are discrete-valued, and this number is likely to keep increasing as chip
complexity rises. Further, the accurate evaluation of a potential design choice
is computationally expensive because it requires detailed cycle-accurate system
simulation.  If the discrete parameter space can be embedded into a larger
continuous parameter space, then continuous space techniques can, in principle,
be applied to the system optimization problem.  Such continuous space
techniques often scale well with the number of parameters.
We propose  a novel technique for embedding the discrete parameter space into
an extended continuous space so that continuous space techniques can be applied
to the embedded problem using cycle accurate simulation for evaluating the
objective function.  This embedding is implemented using simulation-based
ergodic interpolation, which, unlike spatial interpolation, produces the
interpolated value within a single simulation run irrespective of the number of
parameters.  
We have implemented this interpolation scheme in a cycle-based system
simulator. In a characterization study, we observe that the 
interpolated performance curves are continuous, piece-wise smooth, and
have low statistical error.
We use the ergodic interpolation-based approach to solve a large 
multi-core design optimization problem with 31 design parameters.
Our results indicate that continuous space optimization using
ergodic interpolation-based embedding can be a viable approach
for large multi-core design optimization problems.
\footnote{A short version of this paper will be published in the proceedings of IEEE MASCOTS 2015}

\end{abstract}


\section{Introduction}
	
	Modern multi-core systems have complex architectures, containing
	multiple components such as cores, caches and interconnects interacting
	with each other in intricate ways. A system can have tens to hundreds
	of design parameters, most of which are discrete valued (for example,
	size and associativity of caches, latency and throughput of components,
	buffer sizes and issue width of cores), and arriving at a configuration
	that optimizes cost/performance measures such as execution time or
	energy  consumption under given constraints is a non trivial task.
	\emph{Design Space Exploration} (DSE) refers to a systematic process
	for identifying good designs prior to implementation\cite{Gries2004}.
	The set of all possible values that system parameters can take is
	referred to as the \emph{design space} or \emph{parameter space}.  This
	is a multi-dimensional space with each dimension corresponding to a
	design parameter.  Cost/performance measures to be optimized over the
	design space constitute the \emph{objective function}. 
	The optimization process is non-trivial for two reasons:
	\begin{enumerate}
	\item Cost/performance measures cannot be expressed as a function
	of design parameters accurately using simple analytical expressions.
	Simulation of representative benchmark programs on a 
	cycle-accurate model of the system is typically used 
	to evaluate these measures with reasonable accuracy.
	Evaluating each design option is thus computationally expensive.

	\item There are a large number of design parameters.
	The number of possible design configurations grows
	exponentially with the number of dimensions.
	\end{enumerate}

	Techniques for design space exploration aim to find
	good solutions whilst minimizing the computational expense of finding them.
	The computational expense for evaluating a single design option
	is determined by the level of abstraction of the system model chosen.
	Hardware prototypes or FPGA implementations provide performance measures with 
	high accuracy but involve very long implementation time, while  purely 
	analytical models allow faster evaluation but lose out on accuracy. 
	Simulation based evaluation lies between the two extremes.
	Our work focusses on exploring the design space efficiently, 
	assuming that the objective function is evaluated using cycle-accurate simulations.
	Existing techniques for exploring the design 
	space can be broadly classified as follows:
		
	\begin{itemize}
	\item \emph{Exhaustive enumeration:}
	exhaustive search based methods 
	\cite{Givargis1999},
	\cite{JohnsonKin}
	yield globally optimal solutions
	but the number of evaluations becomes prohibitive for large number of parameters. 
	
	\item \emph{Design of experiments (DoE):} number of evaluations can be
	reduced by carefully selecting a subset of points in the design space to be evaluated, using
	design of experiments (DoE) approach
	\cite{Sheldon2007,Yi2003}. However
	effectively using DoE approaches other than full-factorial
	requires prior knowledge about effect of system parameters 
	on performance.

	\item \emph{Search over discrete parameter space:}
	randomized search methods such as simulated annealing \cite{Orsila2008,Schafer2009, Srinivasan1998}, 
	evolutionary algorithms \cite{Holzer2007,Palesi2002,Sengupta2012}
	and heuristic-based local search methods such as hill climbing\cite{Lahiri2000} 
	and Tabu search
	\cite{Eles1997,Wiangtong2002} 
	have been applied to cope with the large dimensionality of the design space. 
	 
	\item \emph{Meta-model based search :} Using systematic sampling,
	a meta model of the system is constructed. The meta-model may be
	in the form of an artificial neural network, linear regression model,
	polynomial or spline interpolation etc. This meta-model is then used 
	in an interleaved manner with simulations to prune the design space
	or guide search during optimization \cite{Palermo2008,Piscitelli2012,Ipek2006}.
	
	\end{itemize}

	\subsection{Main Contributions}

	Existing DSE techniques search for the optimum either
	directly over the discrete parameter space, or search over a meta-model
	which may be defined over a continuous domain.
	We investigate a new approach to this optimization problem which is
	based on \emph{embedding} the discrete parameter space into an extended
	continuous space, and applying continuous optimization
	techniques directly over the embedded simulation model for finding
	local optima efficiently\footnote{Note that this optimization approach
	is distinct from building an interpolation-based meta model of the
	system by sampling the discrete parameter space.}.
	The main motivation behind this approach is to achieve better scalability with
	respect to the number of design parameters. Using continuous optimization
	offers the following advantages:
	\begin{enumerate}
	\item Continuous optimization methods can handle large dimensionality
	of the design space better than exhaustive search or design of
	experiments-based methods.  The number of function evaluations is weakly
	dependent on the number of parameters.

	\item They make use of gradient information and are
	thus more efficient as compared to randomized search
	methods for finding local minima. 

	\item Continuous space offers more pathways to reach 
	the solution as compared to a discrete space. Continuous optimization techniques
	can recognize diagonal ridges in the objective function unlike local search methods in 
	discrete space such as hill climbing \cite{Gries2004}.

	\item The approach does not involve use of a meta-model, thus each function
	evaluation is as accurate as the detailed simulation model.
	\end{enumerate}

	However, descent-based continuous optimization techniques find local
	minima, and random restarts are required to  search for the global
	optimum. Further, in order to convert the continuous space solution
	back to discrete space, rounding needs to be employed with care.   
	The idea of applying continuous optimization techniques to solve a
	discrete optimization problem has been described in the past in
	chemistry\cite{Koh2005} and applied mathematics\cite{Wang2008} .  To
	our knowledge this approach has not been investigated for system-level
	design exploration.

	We propose a technique for embedding discrete parameters into a continuous space 
        by using a simulation-based ergodic interpolation method, which,
	unlike spatial interpolation techniques, can produce the interpolated
	result within a single simulation run irrespective of the number of
	parameters. The basic idea behind ergodic interpolation is to replace
	each discrete parameter in the cycle-accurate model with a discrete
	random variable whose value changes over time, such that the set of
	values of all parameters {\em averaged over time} within a 
	single simulation run approaches a given point in 
	the extended continuous parameter space at which we
	wish to evaluate the interpolated performance value. 
	While ergodic interpolation can be applied, in
	principle, to a variety of discrete parameters, we define and
	demonstrate the embedding for four types of parameters:
	\begin{enumerate}
	\item buffer sizes
	\item component throughputs
	\item component latencies (in units of clock cycles)
	\item number of pipelined stages in interconnect links
	\end{enumerate}
	(where \emph{component} can be a core, cache, memory module 
	or an interconnect).  
	Our primary motivation was to understand the viability of our
	techniques for this relatively simple parameter set.  Based on our
	experience with this parameter set (summarized in Section
	\ref{sec:Results}), we are investigating the application of the ergodic
	interpolation technique to other discrete parameters such as cache size
	and associativity, core issue width etc.
	We first describe the generic system model used in this study
	and define the discrete parameters which are 
	subject to embedding in Section \ref{sec:SystemModel}. 
	We then  describe the ergodic interpolation technique 
	in detail in Section \ref{sec:Embedding}.

	We characterize the interpolated performance function obtained using
	our ergodic interpolation technique on a problem instance with 12
	parameters.  The objective function (total execution time for a
	parallel workload) is evaluated at closely spaced points along random
	straight lines passing through the 12-dimensional continuous parameter
	space. We observe that the interpolated function has low statistical
	error and is continuous and piece-wise smooth along each line.  This
	indicates that continuous space optimization techniques can be used
	with this embedding.  We present the characterization study in Section
	\ref{sec:WellBehavedness}.

	Next, we apply the ergodic interpolation based approach to find 
	optimal configurations in a large DSE problem with 31 discrete parameters.
	The parameters are embedded into a continuous space using ergodic interpolation.
	A variety of continuous space optimization techniques can be applied 
	over this embedding. We choose  COBYLA \cite{COBYLA},
	an algorithm that does not require expensive gradient computations.
	We use an implementation of COBYLA from Python's SciPy library.
	The objective function to be minimized is a weighted sum of performance
	and cost components where weights are varied to obtain cost-performance
	trade-off curves. Performance is defined as the sum of execution times
	for four NAS benchmark kernels and cost is represented by a synthetic
	cost function.  For each set of weights, we perform multiple
	optimization runs starting from random initial points in the design
	space.
	We find that across all optimization runs (given a limit of 300 function evaluations
	per run) the solution converges to a local optimum in all cases and the improvement in 
	objective ranges from 1.3X to 12.2X over the initial guess. Further, the spread 
	in objective values at the optimum across multiple runs is low for most cases.
	We compare the quality of locally optimum solutions found by COBYLA
	runs with a global optimum reported by an Adaptive Simulated Annealing
	(ASA) search in discrete space and find that most of the COBYLA runs
	produce a solution that is close to (within 10\% of) the 
	global optimum reported by ASA.
	The results (presented in Section \ref{sec:Results})
	indicate that continuous space optimization
	applied over an ergodic interpolation based embedding is a viable
	approach for solving discrete optimization problems in design space 
	exploration of multi-core systems.

\section{System Model}\label{sec:SystemModel} \label{sec:SimulationModel}
\begin{figure}[!t]
	\centering
	\includegraphics[width=2.755555in]{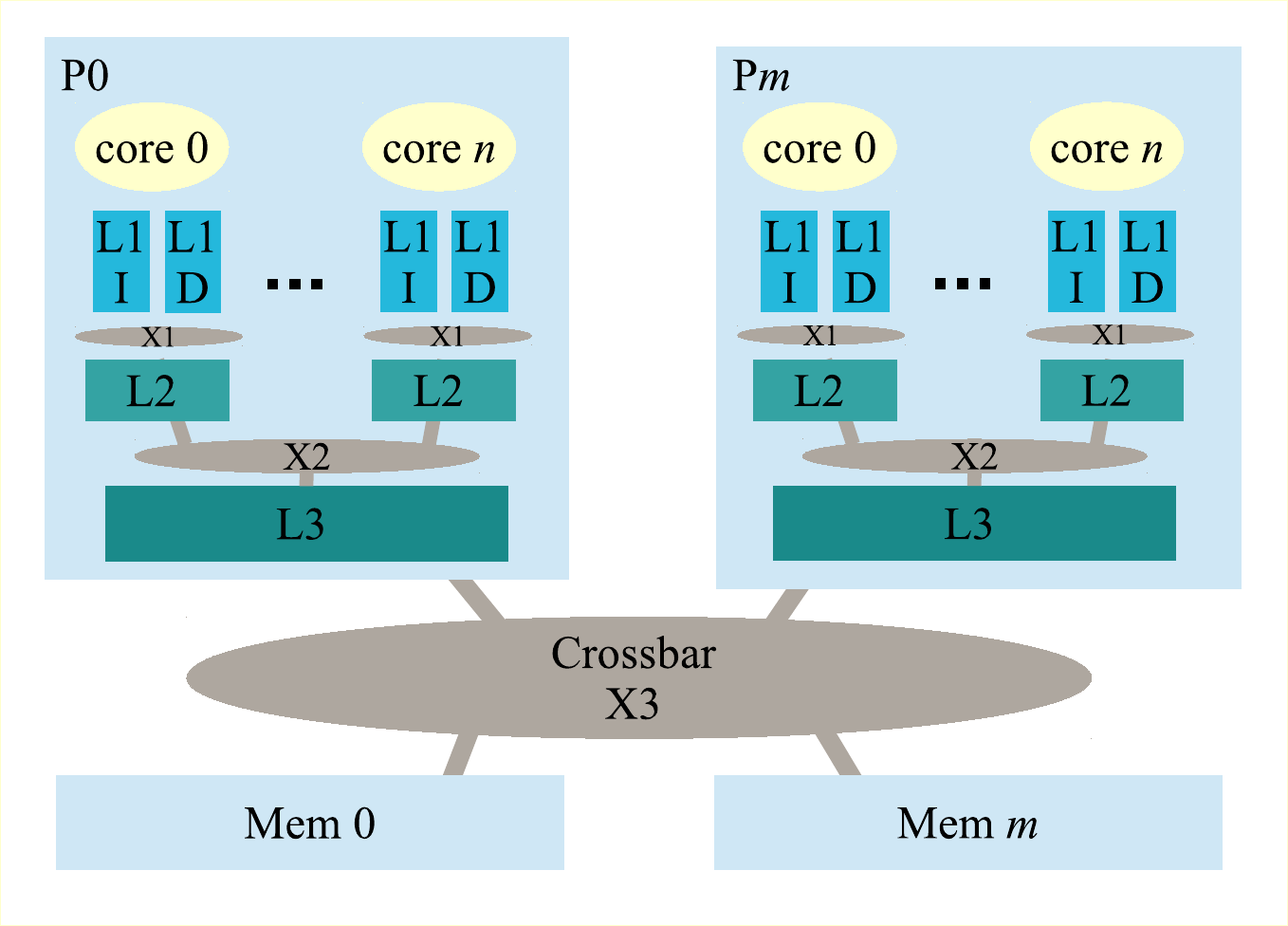}
	\caption{System model}
	\label{fig:SystemModel}
	\end{figure}

	We use a parametrized cycle accurate model of a multi-core system (shown
	in Figure \ref{fig:SystemModel}) that is representative of current NUMA
	(Non Uniform Memory Access) architectures such as those based on the
	Intel QPI\cite{QPI} or AMD HyperTransport\cite{Hypertransport}
	standards.
	The system consists of $m$ processors, with $n$ cores per processor
	(where $m$ and $n$ are model parameters). The processors are connected to 
	$m$ memory modules, forming a $m$-way NUMA configuration.  
	Each core implements the Sparc V8 instruction set.  Timing of
	load/store accesses flowing through the memory subsystem is modeled in
	detail. All other instructions are assumed to execute in one cycle.
	The cache subsystem comprises per-core split L1 and unified L2 caches
	and a shared L3 cache.  Coherency is maintained using a hierarchical
	directory-based MESI protocol which is implemented by generalizing the
	protocol described in \cite[Ch.\ 8.3.2, p.152]{Sorin2011} to an
	arbitrary number of levels in memory hierarchy.	
	Interconnect between successive levels in the memory hierarchy
	is a full-crossbar with parametrized link delays. The NUMA effect is
	modeled by assigning different delays to links connecting a
	processor to its local and remote memory nodes.
	The model is built using the SiTAR modeling framework\cite{sitar}
	and is capable of running user-level programs compiled for
	Sparc V8. Parallel applications can be ported to it using 
	a library of synchronization routines. 
	For all simulations reported in this study,
	the number of processors ($m$) and the 
	number of cores per processor($n$) were fixed at 2 and 4 respectively.

	\subsection{Embedded Parameters} \label{sec:Parameters}

	We present a precise definition of the discrete parameters
	which are subject to embedding.  Although we
	use functional models for all components in the system, each component
	can be classified into one of three basic types:
	\emph{modules}, \emph{wires} and \emph{queues}. Modules represent
	behavioral components in the system such as caches, cores, memory
	modules and interconnect schedulers, wires represent interconnect links
	with a parametrizable number of pipelined stages, and queues are used
	to represent buffering at various places in the system. The system can
	thus be thought of as being constructed using an interconnection of modules (representing the
	processor cores for example), wires (which are pipelined) and buffers/queues.
 	
	%
	%

	The activity in the system (for example, memory accesses and coherence
	requests) is modeled by {\em jobs} and movement of {\em data-tokens}.
	A job represents a behavioural action by a module which can consume and
	produce data-tokens.  Data-tokens are used to encapsulate
	information.  Jobs are triggered inside modules by the availability of
	the necessary data-tokens. The completion of a job may produce new
	data-tokens that can be transported out of the module. Queues are used to
	buffer data-tokens. Each module has its own input and output queues 
	for storing data-tokens to be used by or 
	generated from jobs that it executes. All modules, wires and
	queues in the system are parametrized as follows:

	\begin{itemize}[\setlabelwidth{A}]
	
	\item \textbf{queue capacity $C(q)$ :} A queue $q$ has a single parameter $C(q)$.
	In each cycle, the queue $q$ will accept new data-tokens as long as the total number of
	data-tokens in the queue is $\leq C(q)$.
	
	\item \textbf{module throughput $N(m)$ :} a module $m$ can accept new jobs each cycle as long
	as the number of jobs being processed by it is $\leq N(m)$.

	\item \textbf{module delay $D(j,m)$ :} If $j$ is a job that is accepted by module $m$, 
	then $D(j,m)$ is the number of cycles it takes for the module to execute the job. 
	The job is removed from the input queue as soon as it is accepted by the module and
	place is reserved for its output in the output queues. Its output becomes visible
	in the output queues only after $D(j,m)$ cycles.

	\item \textbf{wire latency $L(w)$ :} A wire $w$ has a single parameter $L(w)$ that represents 
	the number of register stages in the wire. A wire can accept at most one data-token each cycle,
	and each token takes $L(w)$ cycles to pass through the wire. The wire can 
	accept a token only after reserving a place for it in the output queue to
	which it is connected. The token becomes visible 
	in the output queue after $L(w)$ cycles.
	\end{itemize}

	To summarize: our  cycle accurate simulation model has the discrete
	parameters  $C(q)$, $N(m)$, $D(j,m)$ and $L(w)$ 
	for all queues $q$, modules $m$ and wires $w$ that make up the architectural
	components in the system. We define an embedding for these parameters
	in the following section. 	

\section{Embedding the Discrete Parameter Space into Continuous Space}\label{sec:Embedding}
	
		An \emph{embedding} of the discrete parameter space into continuous space
		requires us to extend each discrete parameter to a continuous one. 
		Based on this, we can extend cost/performance functions from the
		discrete space to the extended continuous space by using 
		interpolation.
		The embedding should ideally be implemented 
		in such a way that the behaviour of the interpolated cost/performance functions
		is suitable for the application of continuous optimization algorithms.

	Let $X=\{x_1,x_2,...x_n\}$ be a vector of values of
	discrete-valued design parameters in the model.
	$X\in\Omega_D$
	where $\Omega_D$ is the discrete parameter space. 
	Our cycle based simulation model allows us to evaluate some objective
	function 
	\begin{displaymath}
	f:\Omega_D\rightarrow\mathbb{R}.
	\end{displaymath}
	The function $f$ needs to be optimized.
	We construct an extension of $f$ to produce a continuous function $\hat{f}$:
	\begin{displaymath}
	{\hat{f}}: \Omega_C \rightarrow \mathbb{R}, \ \ \Omega_D \subset \Omega_C \subseteq \mathbb{R}^n
	\end{displaymath}
	where $\Omega_C$  is a continuous space extension of $\Omega_D$.  
	The extension $\hat{f}$ must satisfy
	\begin{equation}
	\hat{f}(Y) = \left\{ \begin{array}{rl}
				f(Y) 		 &	\mbox{ when $Y\in\Omega_D$ } \\
				\theta(Y,\Omega_D) &	\mbox{ otherwise}
	\end{array} \right.
	\end{equation}
	That is, $\hat{f}$ must be continuous in $\Omega_C$ and must agree with $f$ on $\Omega_D$. 
	Thus, we need a suitable \emph{interpolator}  $\theta(Y,\Omega_D)$.

	Spatial interpolation (performed using standard multivariate interpolation methods
	such as Lagrange interpolation\cite{LagrangeInterpolation},
	Simplex interpolation\cite{SimplexInterpolation}
	or Monte Carlo interpolation \cite[p. 143]{MonteCarloInterpolation}
	) is an obvious candidate for $\theta$.
	In spatial interpolation, for each $Y \in \Omega_C$, we identify a set
	of nearest neighbours $X_1(Y),X_2(Y), \ldots X_k(Y)$ of $Y$ such
	that $X_i(Y) \in \Omega_D$ for each $i$.  Then, 
	\begin{displaymath}
	\theta(Y,\Omega_D) = I(X_1(Y),X_2(Y), \ldots X_k(Y))
	\end{displaymath}
	where $I$ is some interpolation function. 
	The interpolated value at a single point $Y$ is then computed 
	in terms of the function values of a set of neighbour points, 
	which have to be computed using expensive simulations.
	Thus, spatial interpolation as a means of embedding is computationally inefficient
	for simulation-based optimization.

	Instead,  we introduce an \emph{ergodic interpolation} method
	which relies on a randomization of the simulation model in
	order to construct the function $\theta$. 
	Using this, the value $\theta(Y)$  can be produced by 
	a single simulation run. The ergodic interpolation method 
	builds on a sensitivity measurement 
	technique described in \cite{Hazari2010}
	for producing small (real-valued) perturbations to 
	discrete-valued parameters in a simulation model.

	\subsection{Ergodic Interpolation using a Randomized Simulation Model}

	The basic idea behind ergodic interpolation is to approximate the 
	result of spatial interpolation with {\em averaging in time}.
	Each discrete parameter $i$ in the model is replaced by a 
	discrete random variable whose value changes over time 
	within a single simulation run, such that its \emph{average} 
	value over the simulation run approaches a real number $v_i \in V$, where 
	the set of average values of all parameters 
	$V = \{v_1, v_2, ... v_n \}$ represents a point in the extended
	continuous space at which we wish to evaluate the interpolated 
	performance value.

	Several choices exist
	in implementing such an embedding: for instance, whether the value
	of the parameter is changed every cycle or once in an interval consisting of 
	multiple cycles, and the exact behavior of a component  when it transitions from 
	one set of values for its parameters to another. We present one possible
	definition of embedding for the  $C(q)$, $N(m)$, $D(j,m)$ and $L(w)$ 
	parameters introduced in Section \ref{sec:Parameters}. We observe that this definition
	leads to interpolated performance functions that are smooth and suitable for continuous space
	optimization.

	In order to construct the ergodic interpolator, we first randomize the
	cycle-based simulation model introduced in Section \ref{sec:SimulationModel}.
	If $0 \leq p \leq 1$ and if $x$ is a real number, then we define
	\begin{displaymath}
	\gamma(p,x) \ = \ \left\{
		\begin{array}{ll}
		\ceil{x} & {\rm with \ probability \ p} \\
		\\
		\floor{x} & {\rm with \ probability \ 1-p}
		\end{array} 
		\right.
	\end{displaymath}
	Thus, $\gamma(p,x)$ is an integer-valued random variable. If $x$ is
	an integer, then $\gamma(p,x) = x$.   For fixed real
	$x$, the expected value of $\gamma(p,x)$ is $p \ceil{x} + (1-p)\floor{x}$.
	It follows that for fixed real $x$, the expected value of $\gamma(x - \floor{x}, x)$
	is $x$.

	Suppose the parameters $C(q), D(j,m), N(m), L(w)$ are real numbers.  Then
	the component behaviour in the simulation model is randomized as follows:
	\begin{itemize}
	
	\item For a queue $q$ with real parameter $C(q)$:   Let $p = C(q) - \floor{C(q)}$.
	At every cycle, accept data-tokens into the queue as long as the total number of tokens
	in the 	queue is $\leq \gamma(C(q) - \floor{C(q)}, C(q))$.
	For example, if $C(q) = 10.3$, then $p = 0.3$.   At each cycle
	$\gamma(C(q)-\floor{C(q)}, C(q))$ will be $11$ with probability $0.3$ and $10$ with
	probability $0.7$, so that during the simulation, the
	queue will have capacity $10$ for 70\% of the time and capacity
	$11$ for the remaining 30\% of the time.
	
	\item For a module with real parameter $N(m)$ : 
	At every cycle, start a new job in the module only if the total number of active jobs
	in the module is less than $\gamma(N(m) - \floor{N(m)}, N(m))$.

	\item For a module $m$ and parameter $D(j,m)$ for some job $j$ : At every cycle,
		if the job $j$ is started successfully, assign a 
		latency of $\gamma(D(j,m) - \floor{D(j,m)}, D(j,m))$ to the job.
	
	\item For a wire $w$ with real parameter $L(w)$ : 
	At every cycle, for a data-token that enters the wire in this cycle, assign a
	transport latency of $\gamma(L(w) - \floor{L(w)}, L(w))$ to the token.
	\end{itemize}

	This randomization effectively ensures that the average value of each 
	parameter can be a real number, while the simulation model continues to
	be  discrete parameter and cycle-based.	  
	The net effect is that each 
	parameter in the simulation model can be treated as a discrete
	valued  Bernoulli random variable whose time-average value is the desired
	continuous value at which the function is to be computed.  
	We call this an ergodic interpolation
	because the time-average in  a single simulation run gives the
	interpolated value.
	The rest of the embedding is easy. We embed $\Omega_D$ into
	a box $\Omega_C$ as follows: for
 	each parameter $p_i$ in the parameter space,
	we define a minimum possible value $m_i$ and a maximum possible value $M_i$.  Then
	\begin{displaymath}
	\Omega_C \ = \ \{ (x_1,x_2,\ldots  x_n) : \ m_i \leq x_i \leq M_i,\ i = 1,2, \ldots n \}
	\end{displaymath}
	For each point $Y \in \Omega_C$, the ergodic interpolation $\theta(Y,\Omega_D)$ for  $Y\in\Omega_C$
	is produced by the randomized simulation model described above.
	This technique gives a well defined interpolation.  However there
	are some questions:
	\begin{enumerate}
	\item What is the amount of statistical error in the interpolated value?
	\item Is the interpolation well-behaved?  That is, is the interpolated
	function smooth enough for us to be able to use continuous optimization
	techniques?
	\end{enumerate}
	We address these questions in the following subsections.

	\subsection{Statistical Error in Ergodic Interpolation}

	Statistical error in the interpolated value can be controlled
	by increasing the number of samples of parameter values.
	This can be done by averaging results 
	from multiple simulation runs, or by using a single long simulation run.
	For the benchmark programs used as workload in our
	design exploration experiment (listed in Table \ref{table:ProblemSizes}),
	we estimated the standard deviation of the interpolated performance value
	at a few points in the design space  by generating multiple samples. 
	For these medium sized benchmarks (spanning 8 to 20 million simulated cycles)
	we find the standard deviation relative to the mean  to be between $0.009\%$ to $0.019\%$.
	These error values are small, and thus, for long enough benchmarks,
	a single simulation run is sufficient for obtaining the interpolated performance value at a single point.

	\subsection{Well-behavedness of Ergodic Interpolation}\label{sec:WellBehavedness}

	We check whether the interpolated performance function
	is smooth, so that continuous optimization techniques can be applied to it.
	We do this by evaluating the function at closely-spaced points 
	along random straight lines passing through the extended continuous 
	parameter space. 
	Parameters for this experiment are $D$, $N$ and $C(\text{ for output
	buffers})$ (as introduced in Section \ref{sec:Parameters}) in L1,
	L2, L3 caches and main memory. Thus the extended continuous 
	parameter space has 12 dimensions.
	The interpolated function $\hat{f}$ is the total time to execute a parallel 
	memory test workload.  The workload involves each core 
	accessing non-overlapping but interleaved memory locations,
	and is chosen to stress the memory system sufficiently.

\begin{figure*}[!ht]
	\captionsetup[subfigure]{labelformat=empty}
	\centering
	\subfloat[]{\includegraphics[width=3.0in]{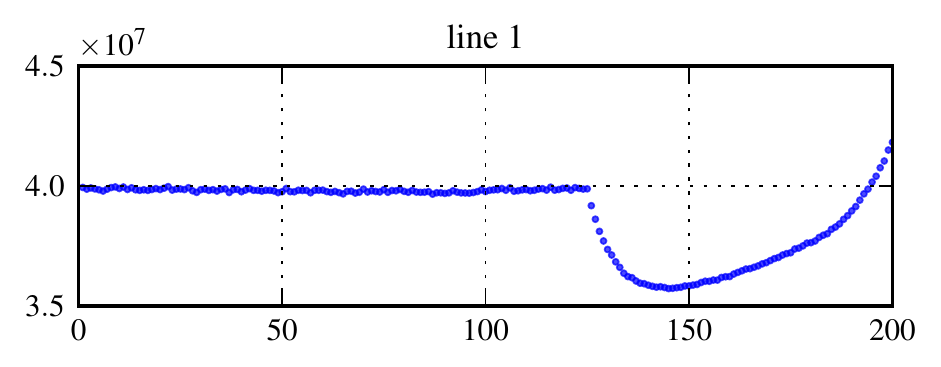}}%
	\hfil
	\subfloat[]{\includegraphics[width=3.0in]{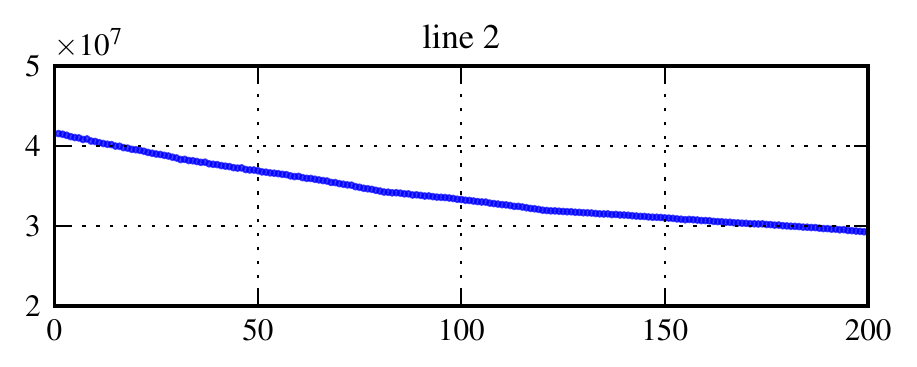}}%
	\vspace{-1.5em}
	\subfloat[]{\includegraphics[width=3.0in]{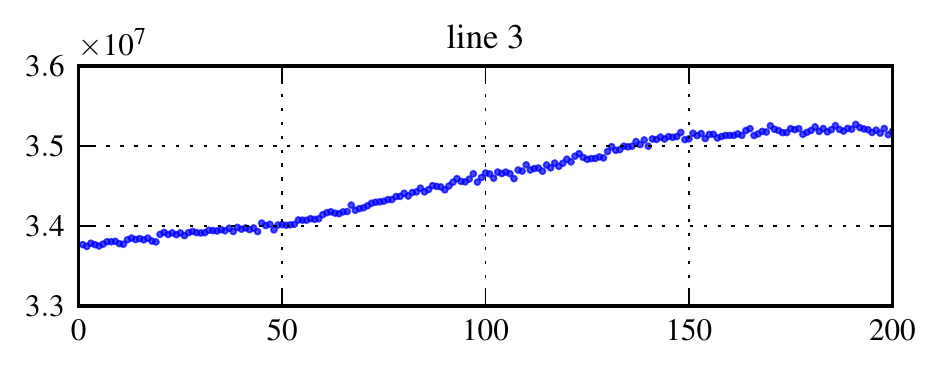}}%
	\hfil
	\subfloat[]{\includegraphics[width=3.0in]{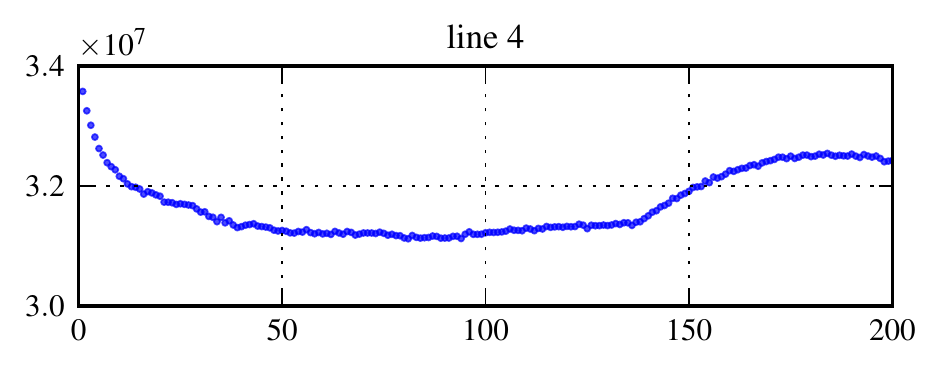}}%
	\vspace{-1.5em}
	\subfloat[]{\includegraphics[width=3.0in]{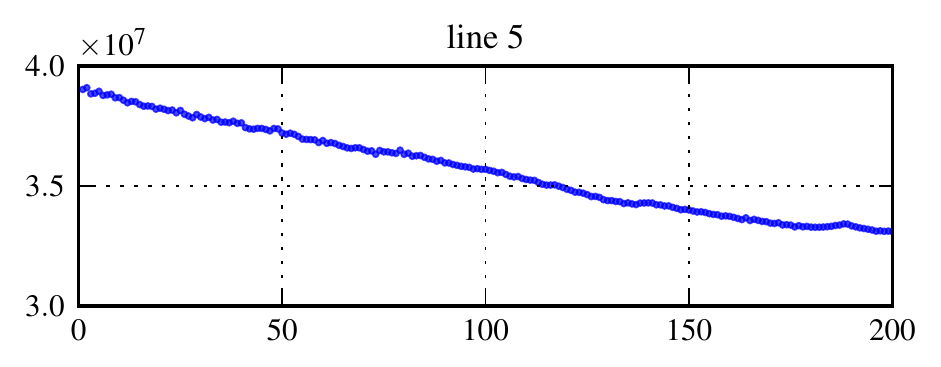}}%
	\hfil
	\subfloat[]{\includegraphics[width=3.0in]{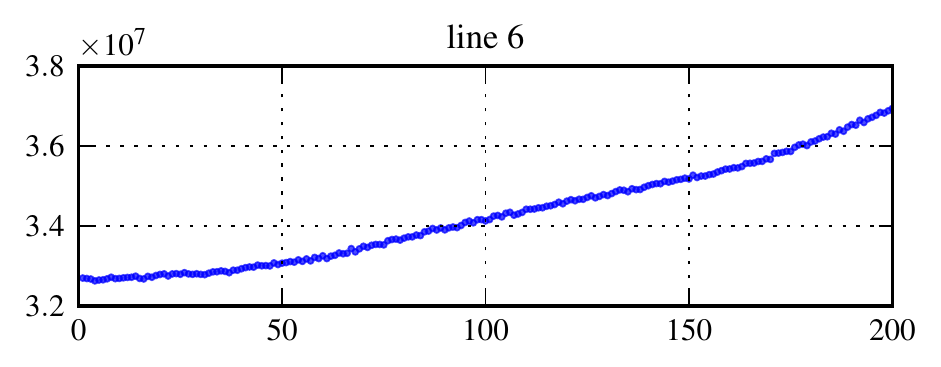}}%
	\vspace{-1.5em}
	\subfloat[]{\includegraphics[width=3.0in]{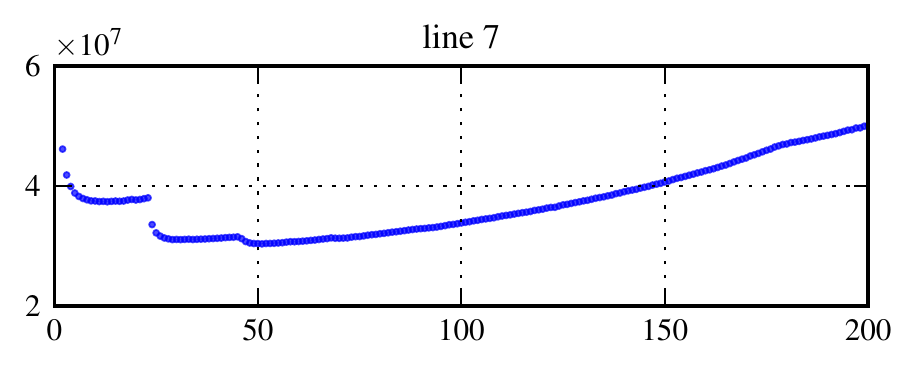}}%
	\hfil
	\subfloat[]{\includegraphics[width=3.0in]{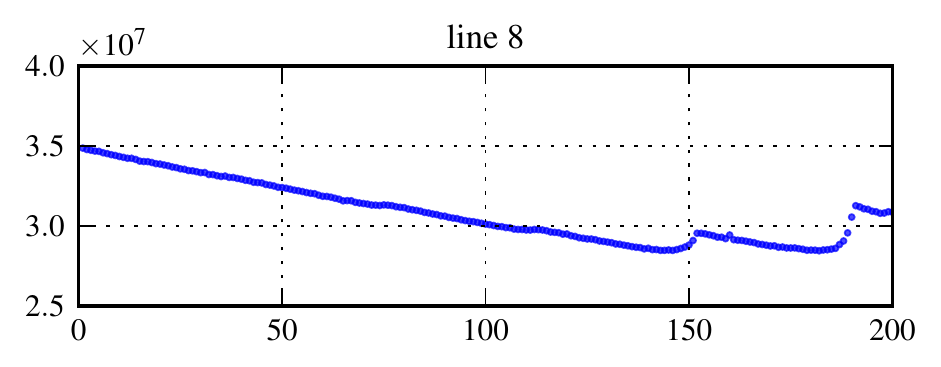}}%
	\vspace{-1.5em}
	\subfloat[]{\includegraphics[width=3.0in]{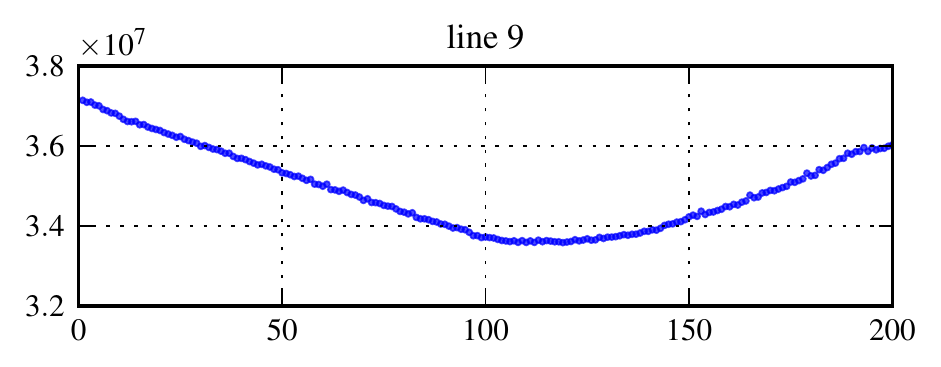}}%
	\hfil
	\subfloat[]{\includegraphics[width=3.0in]{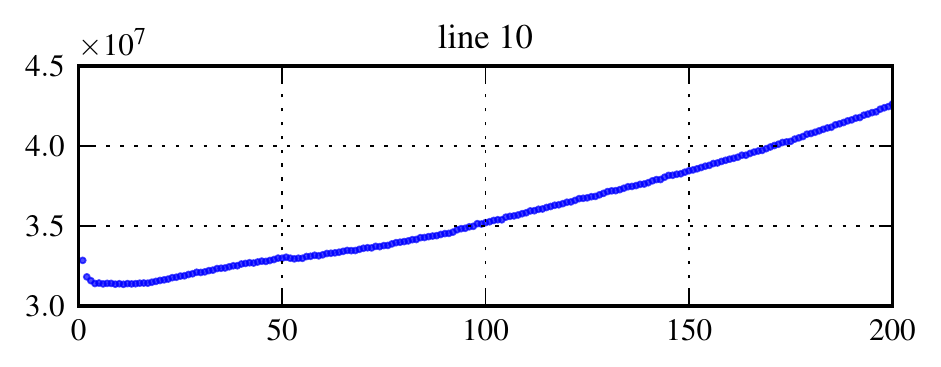}}%
	
	\caption{Interpolated  objective function (total execution time) values 
	plotted along ten random straight lines passing through the multi-dimensional
	continuous parameter space. Each line is sampled at 200 uniformly 
	spaced points.}
	\label{fig:WellBehavedPlots}
	\end{figure*}

	We consider random straight lines passing through the 
	12-dimensional continuous parameter space. Each line is sampled at
	200 uniformly spaced points, and the objective function 
	$\hat{f}$ (total execution time) is evaluated at each of these points
	using cycle accurate simulations of the model described in Section \ref{sec:SystemModel}. 
	We perform multiple simulation runs at each point with distinct randomization seeds 
	in order to measure the mean and standard error values.
	In Figure \ref{fig:WellBehavedPlots},  we show the interpolated performance function values
	(mean $\hat{f}$) evaluated at uniformly spaced points along ten randomly chosen 
	straight lines passing through the continuous parameter space.
	We observe that along each line, the interpolated function is continuous and piece-wise smooth.
	The measured relative standard error values are less than $0.01\%$.  
	Thus the interpolated function obtained using our ergodic interpolation technique
	seems to be well-behaved and suitable for the application of 
	continuous optimization techniques.

\section{Results of Continuous Optimization using Ergodic Interpolation}\label{sec:Results}
	
	We apply the ergodic interpolation based approach for 
	finding optimal configurations in a multiprocessor design exploration
	problem with 31 discrete parameters. The discrete parameters and their 
	ranges are listed in Table \ref{table:DesignParameters}. The parameters
	are embedded into continuous space using ergodic interpolation over a
	cycle accurate model described in Section \ref{sec:SystemModel}.
	Four kernels from the NAS parallel benchmark suite
	(NPB)\cite{NASbenchmarks} are used as workload.  We have ported an
	OpenMP+C version of NPB v2.3 developed by the Omni
	project\cite{OmniNAS} to our model. The kernels and their problem sizes
	are listed in Table \ref{table:ProblemSizes}.

	\begin{table}[!htb]
	\renewcommand{\arraystretch}{1.22}
	\caption{NAS kernels and their problem sizes}
	\label{table:ProblemSizes}
	\centering
	\begin{tabular}{| c | c |}
	    \hline
	    Kernel & Problem Size\\ \hline\hline
	    Embarrassingly Parallel (EP) & $2  ^ {16}$	\\\hline
	    Multigrid (MG)		 & $16 ^ 3$	\\\hline
	    3-D FFT PDE solver (FT)	 & $16 ^ 3$	\\\hline
	    Integer Sort (IS)		 & $2  ^ {16}$	\\\hline
	    \end{tabular}
	\end{table}

	\setlength{\tabcolsep}{4pt}
	\begin{table}[!tb]
	\renewcommand{\arraystretch}{1.2}
	\caption{Design Parameters and their Ranges\newline
	('Opt' lists parameter values at the optimum for $\alpha=10^5$)}

	\label{table:DesignParameters}
	\centering
	\begin{tabular}{| l | c | c | r || l | c | c | r |}
	    \hline
	Parameter			&	Min		&	Max	&	{Opt}	&	 Parameter			&	Min		&	Max	&	{Opt}		\\ \hline\hline
	$N$(L1I)			&	1		&	4	&	{2.95}	&	 $C_{\text{inQ}}$(L1I)		&	1		&	4	&	{1.00}		\\ \hline		
	$N$(L1D)			&	1		&	4	&	{1.93}	&	 $C_{\text{inQ}}$(L1D)		&	1		&	4	&	{1.06}		\\ \hline		
	$N$(L2)				&	1		&	4	&	{1.27}	&	 $C_{\text{inQ}}$(L2)	 	&	1		&	16	&	{1.12}		\\ \hline		
	$N$(L3)				&	1		&	4	&	{1.22}	&	 $C_{\text{inQ}}$(L3)		&	1		&	16	&	{2.22}		\\ \hline		
	$N$(mem)			&	1		&	4	&	{1.02}	&	 $C_{\text{inQ}}$(mem)		&	1		&	32	&	{1.00}		\\ \hline		
	$D$(L1I)			&	1		&	4	&	{ 1.70}	&	 $C_{\text{outQ}}$(L1I)		&	1		&	4	&	{1.06}		\\ \hline 	
	$D$(L1D)			&	1		&	4	&	{ 3.17}	&	 $C_{\text{outQ}}$(L1D)		&	1		&	4	&	{1.00}		\\ \hline 	
	$D$(L2)				&	8		&	16	&	{ 9.33}	&	 $C_{\text{outQ}}$(L2)		&	2		&	16	&	{2.00}		\\ \hline 	
	$D$(L3)				&	16		&	32	&	{21.90}	&	 $C_{\text{outQ}}$(L3)		&	4		&	16	&	{4.00}		\\ \hline 	
	$D$(mem)			&	64		&	128	&	{80.61}	&	 $C_{\text{outQ}}$(mem)		&	4		&	32	&	{4.00}		\\ \hline		
	$L$(X1)				&	1		&	4	&	{2.18}	&	 $C_{\text{inQ}}$(X3)		&	1		&	8	&	{1.00}		\\ \hline
	$C_{\text{inQ}}$(X1)		&	1		&	4	&	{1.00}	&	 $L$(X3) local			&	16		&	64	&	{62.53}		\\ \hline
	$C_{\text{outQ}}$(X1)		&	1		&	4	&	{1.00}	&	 $L$(X3) remote			&	32		&	64	&	{55.27}		\\ \hline
	$L$(X2)				&	4		&	8	&	{5.41}	&	 $C_{\text{outQ}}$(X3) local	&	1		&	16	&	{1.00}		\\ \hline
	$C_{\text{inQ}}$(X2)		&	1		&	4	&	{1.00}	&	 $C_{\text{outQ}}$(X3) remote	&	1		&	16	&	{1.02}		\\ \hline
	$C_{\text{outQ}}$(X2)		&	1		&	4	&	{1.02}	&					&			&		&	{}		\\ \hline
	\multicolumn{8}{|l|}{
	\parbox[t]{3in}{
	\begin{itemize}
	\item $N$, $D$, $L$ and $C$ are parameters as described in Section \ref{sec:Parameters}. 
	\item Components L1I, L1D, L2 and L3 are caches as depicted in Figure \ref{fig:SystemModel}, mem refers to a main memory bank,
	and X1, X2, and X3 refer to interconnects between L1 to L2, L2 to L3 and L3 to main memory respectively.
	\item Links in X3 connecting a processor to its local and remote memory modules have different $L$ and $C_{\text{outQ}}$ values to model NUMA effect.
	\end{itemize}
	}}\\ \hline 
	\end{tabular}
	\end{table}

	Continuous optimization is performed over the embedding
	using an implementation of a derivative-free continuous optimization 
	algorithm COBYLA\cite{COBYLA} from Python's SciPy library.
	We define the objective function for this optimization experiment
	as follows:

	\subsection{The Objective Function}\label{sec:ObjectiveFunction}

	Let $Y \in \mathbb{R}^n$ be a vector of parameter 
	values in the extended continuous space and $\hat{f}(Y)$
	denote the objective function we wish to minimize.
	We construct the objective function as a weighted sum of
	performance and cost measures.
	The performance measure ${\text{execution\_time}}(Y)$ is the sum of
	execution times for four benchmark kernels (listed in Table
	\ref{table:ProblemSizes}).

	We represent cost using a synthetic function ${\text{cost}}(Y)$ 
	which increases as each parameter is varied in the direction of improving performance.
	The cost function is defined as ${\text{cost}}(Y) = \sum_i{x_i} + \sum_j{\frac{100}{d_j}}$
	where $d_j$ and $x_i$ are delay and non-delay parameters
	normalized to lie in the range $[1,100]$.
	This synthetic cost function is sufficient to demonstrate the validity
	of our technique and can be replaced with other cost functions 
	such as energy consumption as appropriate for a particular design
	exploration problem. The objective function is:
	$$
	\hat{f}(Y) = {\text{execution\_time}} (Y) + \alpha \times {\text{cost}} (Y)
	$$
	Where $\alpha$ is a weighting factor which is varied to obtain cost/performance trade offs.

	\subsection{Results}

        We study the convergence properties of 
	the COBYLA algorithm applied to the objective function
	obtained by our ergodic interpolation technique.
	Optimization is performed with multiple values of the weight factor
	$\alpha\in\{0,10^4,10^5,10^6\}$ to get cost/performance trade-off
	curves.  Further, we perform eight optimization runs starting from distinct
	randomly chosen points in the parameter space for each value of $\alpha$.
	A single optimization run is allowed to make at most 300 function evaluations.

\begin{figure}[!t]
	\centering
	\includegraphics[width=3.2in]{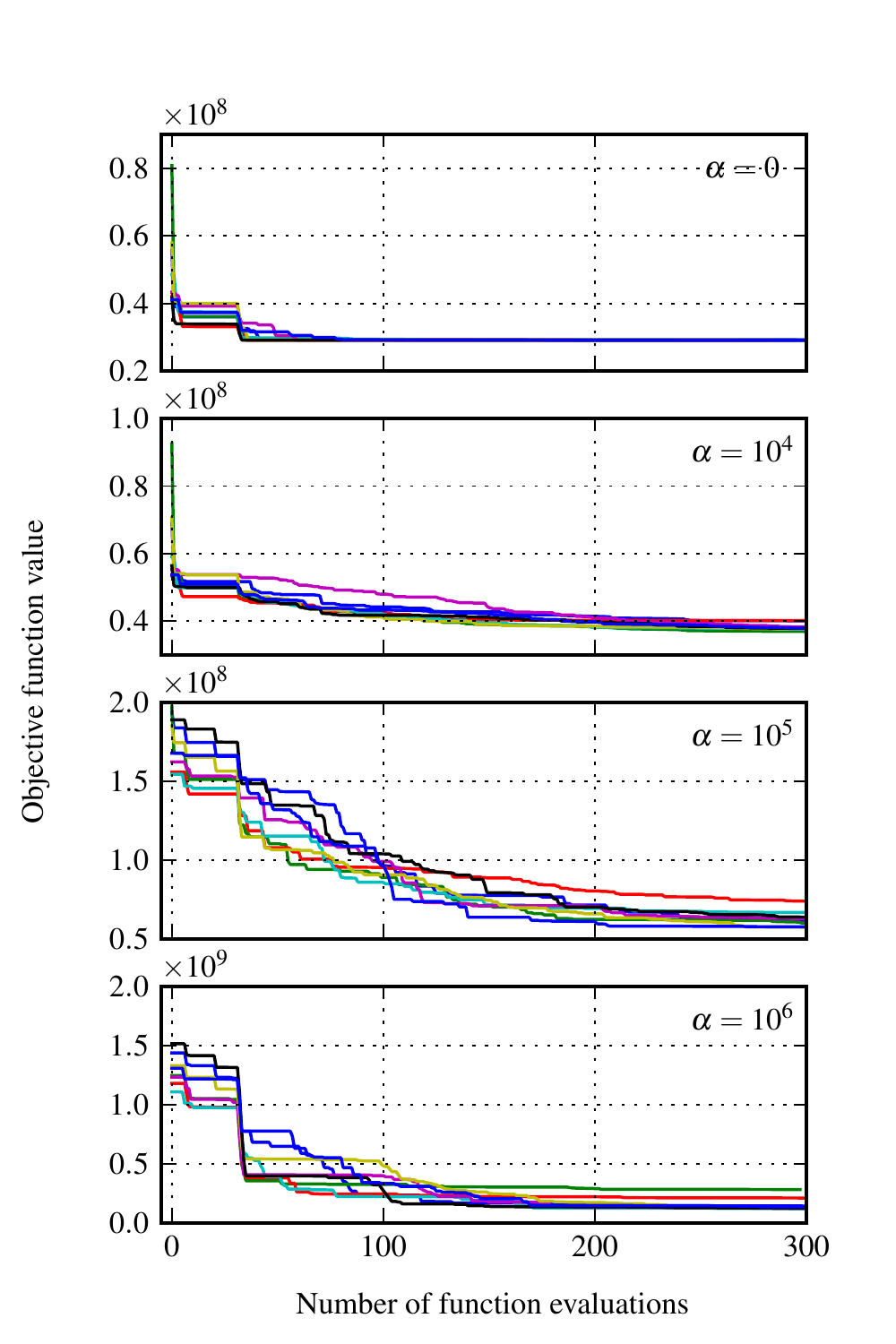}
	\caption{Objective function values versus number of function evaluations.
	For each value of the weighting factor $\alpha$, eight optimization runs are performed
	starting from distinct initial points.}
	\label{fig:ConvergenceRate}
	\end{figure}

	In Figure \ref{fig:ConvergenceRate}, we show the evolution of the
	objective function with the number of function evaluations for all optimization runs.
	In almost all cases, we observe that the objective function
	values converge to a local optimum within 200 function
	evaluations.  Further, the spread of values obtained
	across different initial points is small for most cases as summarized in 
	Table \ref{table:ObjectiveValueSpread}. Improvements over the initial guess
	range from 1.3X to 12.2X as listed in Table \ref{tab:Improvements}.

	\begin{table}[!htb]
	\renewcommand{\arraystretch}{1.3}
	\caption{Objective function values at the optimum across eight COBYLA runs for each value of $\alpha$}
	\label{table:ObjectiveValueSpread}
	\centering
	\begin{tabular}{|l|c|c|c|c|}
		\hline
		                & $\alpha=0$		& $\alpha=10^4$		& $\alpha=10^5$		& $\alpha=10^6$		\\\hline \hline
		best		& $2.916\times 10^7$	& $3.697\times 10^7$	& $5.753\times 10^7$	& $1.218\times 10^8$	\\\hline	
		worst		& $2.922\times 10^7$	& $4.020\times 10^7$	& $7.401\times 10^7$	& $2.836\times 10^8$	\\\hline
		mean		& $2.917\times 10^7$	& $3.836\times 10^7$	& $6.305\times 10^7$	& $1.608\times 10^8$	\\\hline
		relative std dev		& $0.08\%$ 		& $3.08\%$		& $8.59\%$		& $35.66\%$		\\\hline
	\end{tabular}
	\end{table}

	\begin{table}[!htb]
	\renewcommand{\arraystretch}{1.3}
	\caption{Improvements in objective function value over the initial guess}
	\label{tab:Improvements}
	\centering
	\begin{tabular}{| c | r | r | r | r |}
		\hline
		\textbf{ } &\textbf{$\alpha=0$} &\textbf{$\alpha=10^4$} & \textbf{$\alpha=10^5$} &\textbf{$\alpha=10^6$}\\\hline \hline
		best  & 2.8x & 2.5x & 3.3x & 12.2x \\\hline	
		worst & 1.4x & 1.3x & 2.1x & 4.4x  \\\hline
		mean  & 1.8x & 1.7x & 2.8x & 8.8x  \\\hline
	\end{tabular}
	\end{table}

\begin{figure}[!tb]
	\centering
	\includegraphics[width=3in]{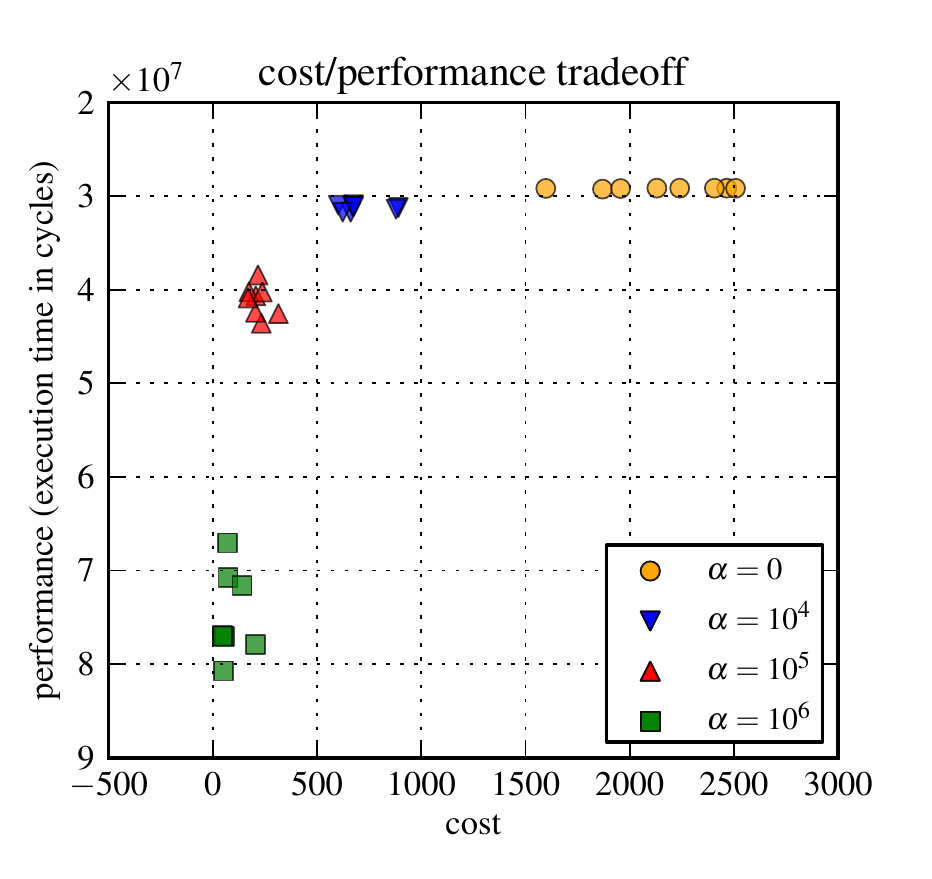}
	\caption{Performance and cost values at the optimum for multiple optimization runs
	(as $\alpha$ and initial points are varied).
	Each point represents the result of a single optimization run with (x,y) coordinates
	showing (cost, performance) values at the optimum. }
	\label{fig:CostPerformance}
	\end{figure}

	In Figure \ref{fig:CostPerformance}, we plot cost and performance values
	at the optimum for each of the optimization runs
	(as $\alpha$ and initial points are varied). Each point in the plot
	represents the result of a single optimization run with (x,y) coordinates
	showing (cost, performance) values at the optimum. The plot shows a clear {\em knee}
	which can be used to select the optimal system configuration for maximum performance.
	Parameter values at the best solution among all COBYLA runs (for $\alpha=10^5$ at the knee) 
	are listed in Table \ref{table:DesignParameters}.  The solution indicated in 
	Table \ref{table:DesignParameters} is interesting.  For example, it indicates
	that the throughput parameter $N$ for the L1 I-cache should be approximately
	3, while that for the L1 D-cache should be approximately 2.  Further, the
	delay $D$ of the L1 D-cache can be considerably larger than that of the
	L1 I-cache (3.17 versus 1.70).

	Since the COBYLA algorithm produces a local optimum, we are also
  	interested in understanding the quality of the local optimum
	across random initial starting points.
	We compare the quality of solutions generated by COBYLA to those
	generated by an Adaptive Simulated Annealing (ASA) search over  
	discrete parameter space. We use a Python binding\cite{pyasa}
	of a well-established ASA implementation \cite{Ingber1996}.
	In Table \ref{table:ASA}, we show the objective function values
	obtained after running ASA with a limit of 1000 function evaluations.
	We observe that most of the COBYLA runs produce solutions that are
	close to (within 10\% of) the global optimum reported by ASA 
	as summarized in Table \ref{tab:SolutionsCloseToASA}.

	\begin{table}[!htb]
	\renewcommand{\arraystretch}{1.3}
	\caption{Objective function values at the optimum reported by ASA}
	\label{table:ASA}
	\centering
	\begin{tabular}{|c|c|c|c|} 
		\hline
		$\alpha=0$		& $\alpha=10^4$		& $\alpha=10^5$		& $\alpha=10^6$ 	\\\hline \hline
		$2.916\times10^7$	& $3.694\times10^7$	& $6.063\times10^7$	& $1.247\times 10^8$	\\\hline
	\end{tabular}
	\end{table}

	\begin{table}[!htb]
	\renewcommand{\arraystretch}{1.3}
	\caption{Number of COBYLA runs (out of 8) that yield solutions within 10\% of
	the global optimum reported by ASA}
	\label{tab:SolutionsCloseToASA}
	\centering
	\begin{tabular}{|c|c|c|c|} 
		\hline
		$\alpha=0$	& $\alpha=10^4$	& $\alpha=10^5$	& $\alpha=10^6$ 	\\\hline \hline
		8		& 8		& 6		& 5	\\\hline
	\end{tabular}
	\end{table}

\section{Conclusions}\label{sec:Conclusions}

	We have described a technique using which discrete parameter
	multi-core systems can be optimized using continuous space
	optimization schemes.  The technique relies on a novel ergodic
	interpolation scheme based on randomizing the discrete parameter
	cycle-accurate simulation model of the multi-core system.
	The interpolated performance function is continuous, 
	has low statistical error, and was observed 
	to be piece-wise smooth.
	Using this ergodic interpolation technique,
	we have applied a standard continuous space optimization algorithm 
	to find optimal designs for a 31-parameter multiprocessor
	system exercised with a subset of the NAS benchmarks.  The 
	optimization algorithm converged to a local optimum within 
	200 function evaluations and produced substantial improvements 
	ranging from 1.3X to 12.2X over the initial 
	guess in the cases that we have tried.  
	Cost performance curves can also be generated using different weightings of the performance
	and cost components in the objective function.  

        More work is needed to completely characterize the impact of rounding 
	on the quality of the results obtained,  and on the application of
	the ergodic interpolation technique to other discrete parameters such as cache size/associativity and
	processor core issue-width/clock-frequency.  However, our preliminary
 	investigations indicate that ergodic interpolation based optimization can be
	an effective and practical approach for the design space exploration of 
	multi-core systems.

\section*{Acknowledgment}
Part of this work was funded by an IBM Faculty Award.
Simulations in Section \ref{sec:WellBehavedness} were run
on CDAC's PARAM Yuva II cluster.
The authors wish to thank Prof. Virendra Sule for granting us
the use of a 48 core cluster and Prof. Sachin Sapatnekar for 
his suggestions and feedback during initial stages of this work.

\bibliographystyle{IEEEtran}
\bibliography{IEEEabrv,bibliography_list}
%

%
%

\end{document}